\documentclass[twocolumn,amsmath,amssymb]{snp}
\usepackage{graphicx}
\usepackage{dcolumn}
\usepackage{bm}
\begin{document}

\title{\Large Spectator matter fragmentation in Au+Au reactions: Phase space analysis}

\author{\large Yogesh K. Vermani}
\email{yugs80@gmail.com}
\affiliation{Department of Applied Science and Humanities, ITM University, Gurgaon-122017,
INDIA}


\maketitle

\noindent {\large{\bf Introduction}}

The reacion mechanism behind  multi fragmentation in heavy-ion collisions has been a long standing question 
in intermediate energy nuclear physics. Not only the multiplicities of various kinds of fragments are useful 
observables, but how and when these fragments are realized in phase space can also shed light on reaction dynamics. 
These correlations can be of importance to reveal the phase space details of decaying system. In this direction, 
Puri and collaborators have devised a faster cluster recognition algorithm namely \textit{simulated annealing clusterization algorithm} 
(SACA) \cite{jcp, epl, dae}. This algorighm is found particularly successful in describing spectator matter break-up at relativistic 
\cite{epl, jpg} and ultra low \cite{cejp} incident energies. The phenomenon of global universality of \textit{rise and fall} pattern in the multiplity of intermediate mass fragments was accuratley reproduced with this model \cite{epl}. 
It, however, remained to see how these fragments are correlated in phase space and originate from which region of the nucleus. 
In the present paper, we shall compare the phase space characteristics of the intermediate mass fragments (IMFs) produced using \textit{minimum spanning tree} (MST) and SACA clusterization algorithms. \\

\noindent {\large{\bf The Model}}

The phase space of nucleons is recorded at different time steps 
using \textit{quantum molecular dynamics} (QMD) transport theory \cite{hart}. In QMD approach, 
the trajectories of nucleons follow Hamilton's equations of motion:
\begin{eqnarray}
\dot{{\bf p}_i}=& -\{{\bf r}_i, {\cal H}\}, \\
\dot{{\bf r}_i}=&\{{\bf p}_i, {\cal H}\}.
\end{eqnarray}
Here ${\cal H}$ is the total Hamiltonian of the system of
$A_{P}+A_{T}$ nucleons. These equations are solved after fixed
time interval $ \bigtriangleup t$ chosen to be very small. The phase space of nucleons is clusterized using MST and SACA methods. 
In SACA model, pre-clusters obtained with minimum spanning tree procedure are subjected to a binding energy cut \cite{jcp,epl}:
\begin{eqnarray}
\zeta_{a}=\frac{1}{N_{f}}\sum_{i=1}^{N_{f}}\left[\sqrt{\left(\textbf{p}_{i}-\textbf{P}_{N_{f}}^{cm}\right)^{2}+m_{i}^{2}}-m_{i}
+ \right . \nonumber \\
\left. \frac{1}{2}\sum_{j\neq i}^{N_{f}}V_{ij}
\left(\textbf{r}_{i},\textbf{r}_{j}\right)\right]< -E_{bind},
\label{be}
\end{eqnarray}
with $E_{bind}$ = 4.0 MeV if $N_{f}\geq3$, else $E_{bind} = 0$. In
this equation, $E_{bind}$ is the fragment's binding energy per
nucleon, $N_{f}$ is the number of nucleons in a fragment, and
$P_{N_{f}}^{cm}$ is the center-of-mass momentum of the fragment. \\


\noindent {\large{\bf Illustrative Results}}

\begin{figure}
\includegraphics [scale=0.8]{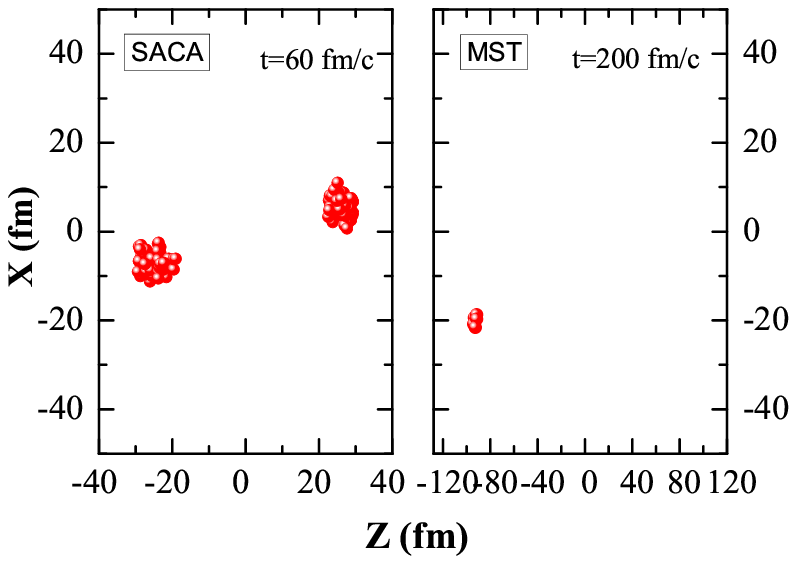}%
\vskip -0.3cm 
\caption {The distribution of nucleons bound in IMFs is displayed in the reaction plane (\textit{i.e.} Z-X plane) 
using SACA (left) and MST (right) methods.}
\vskip 1.6cm
\includegraphics [scale=0.8]{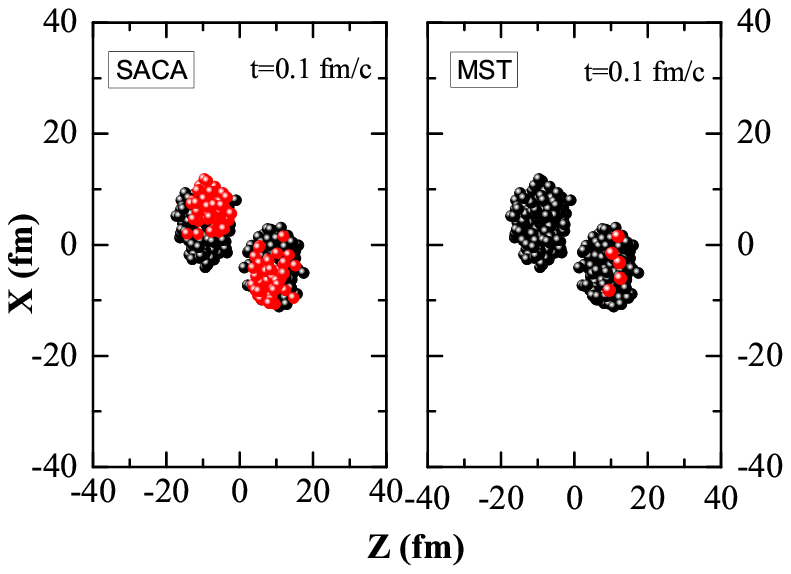}%
\vskip -0.3cm
\caption {Nucleons distribution in the two colliding Au-nuclei in the Z-X  plane at the time of initial contact. 
Filled circles in red indicate positions of nucleons emitted as intermediate mass fragments.} 
\vskip 0.4cm 
\end{figure}

For this study we simulate a single event of $^{197}$Au+ $^{197}$Au collision at incident 
energy $E_{lab}=$ 1000 AMeV and impact parameter $b=$ 8 fm. We have employed a soft EoS (${\cal K}$=200 MeV) along 
with energy dependent \textit{n-n} cross section. We display in Fig. 1 the distribution of nucleons bound in IMFs
[$5\leq A \leq 65$] using SACA (at 60 fm/c) and MST (at 200 fm/c) methods. 
One can see that higher yield of fragments is obtained using SACA method at forward as well as backward rapidities. 
MST approach, on the other hand, predicts only single cluster of mass $A=5$ in the backward hemi-sphere. This shows that SACA method is much 
faster in recognizing the proper fragment yields from the spectator matter break-up.

Next, we backtrack the original location of these nuclei in the colliding nuclei at the start of nucleus-nucleus collision. 
Figure 2 shows these nucleons emanating as bound clusters in the two Au-nuclei. One can see that SACA 
method leads to exaggerated yield of fragments out of spectator domain. In MST method, however, nucleons are confined to overlapped zone only. 
This results in under-estimated yield of fragment multiplicities in ALADiN data.

Summarizing, we addressed in this paper the problem of origin of fragments via phase space analysis of MST and SACA clusters. Our model predictions highlights the importance of binding energy criterion in recognition of most bound fragment configuration. 
This criterion is important to understand the spectator matter physics and accurately predicting fragment multiplicities. \\

\vfill
\end{document}